\documentclass[AMA,STIX1COL]{WileyNJD-v2}
\usepackage{moreverb}
\usepackage{cleveref}

\newcommand\BibTeX{{\rmfamily B\kern-.05em \textsc{i\kern-.025em b}\kern-.08em
T\kern-.1667em\lower.7ex\hbox{E}\kern-.125emX}}

\articletype{Special Issue}%

\received{<day> <Month>, <year>}
\revised{<day> <Month>, <year>}
\accepted{<day> <Month>, <year>}

\begin{document}

\title{LLMs for Science: Usage for Code Generation and Data Analysis}

\author[1]{Mohamed Nejjar}

\author[1]{Luca Zacharias}

\author[1]{Fabian Stiehle}

\author[1,2]{Ingo Weber}

\authormark{Nejjar and Zacharias \textsc{et al.}}
\titlemark{LLMs for Science}

\address[1]{\orgname{Technical University of Munich}, \orgdiv{School of Computation, Information and Technology}, \orgaddress{\city{Munich}, \country{Germany}}}

\address[2]{\orgname{Fraunhofer Gesellschaft}, \orgaddress{\city{Munich}, \country{Germany}}}

\corres{*Fabian Stiehle, Technical University of Munich, School of Computation, Information and Technology\email{fabian.stiehle@tum.de}}

\abstract[Abstract]{Large language models (LLMs) have been touted to enable increased productivity in many areas of today's work life. Scientific research as an area of work is no exception: the potential of LLM-based tools to assist in the daily work of scientists has become a highly discussed topic across disciplines. However,
we are only at the very onset of this subject of study. It is still unclear how the potential of LLMs will materialise in research practice. With this study,
we give first empirical evidence on the use of LLMs in the research process.
We have investigated a set of use cases for LLM-based tools in scientific research, and conducted a first study to assess to which degree current tools are helpful.
In this paper we report specifically on use cases related to software engineering, such as generating application code and developing scripts for data analytics. While we studied seemingly simple use cases, results across tools differ significantly. Our results highlight the promise of LLM-based tools in general, yet we also observe various issues, particularly regarding the integrity of the output these tools provide.}


\maketitle

\section{Introduction and Background}
\label{sec:introduction}


With the public release of \textit{ChatGPT} in November 2022, large language models (LLMs) attracted widespread public attention. LLMs are machine learning models, often based on neural networks following a pre-trained transformer architecture, which have many parameters and have been trained on a large corpus of training data.\cite{MinLLMSurvey} At the time of writing, hundreds of billions of model parameters are not a rare occurrence, as is training data with a trillion tokens.\footnote{\url{https://en.wikipedia.org/wiki/Large_language_model\#List}, accessed 2023-11-03}
Besides various other use cases, the ability of LLMs to generate meaningful textual answers to natural-language text queries has captivated the imagination of many, with productivity use cases at the forefront. The scientific research community is no different here.
In their global survey on the work of postdocs, \textit{Nature} found that 31\% of postdocs are already using AI-assisted chatbots like ChatGPT. Of those who use it, the two most reported use cases are text refinement and generating or troubleshooting code.\cite{le2023chatgpt} Among disciplines, the potential (and dangers) of LLM-based tools has quickly become a highly discussed topic, and has given rise to many  editorial, opinion, and commentary pieces, with most expecting high productivity gains as the most promising outcome of utilising these tools.\cite{Susarla:2023,li2023ethics,grossmann2023ai,kasneci2023chatgpt}

Scientific work comprises many text-based tasks, such as reading and synthesizing publications, writing laboratory logs, creating code for software prototypes, data analysis, writing papers, and creating slide decks for presentations. A recent paper from the information systems discipline further lists tasks like problem formulation, research design, and data collection.\cite{Susarla:2023} All of these tasks could in principle be supported by LLM-based tools. We refer to this direction of research as \textit{LLMs4Science}.
LLMs are part of a new generation of artificial intelligence (AI) artifacts and methods, called \textit{generative AI}, which further comprises AI that is able to generate images, videos, speech, music, and more.
For the use of generative AI in science, we propose the term \textit{GenAI4Science}.
However, the use of such tools is subject to many---not yet well understood---risks. One major problem is commonly referred to as \textit{``hallucination''}---though a more precise term is \textit{``confabulation''}\cite{Smith:2023:Confabulation}---where an LLM seemingly fills gaps in the information contained in the model with plausibly-sounding words. 
Given their mode of operation, i.e., predicting the most likely next word from a neural network that has commonly been trained with a large corpus of text, the occurrence of this effect can be explained well.
For application areas like scientific writing, this causes a significant issue: factual correctness is of utmost importance, and writing that does not adhere to the rigorous standards of scientific integrity should not be considered part of the body of scientific knowledge (for a negative example of a confabulating LLM in science, see \textit{Galactica} in Section~\ref{sec:related}).

Despite problems, challenges, and past setbacks, the potential of LLMs to achieve increased productivity in science appears very high. However, we are only at the very onset of this subject of study. It is still unclear how the potential of LLMs will materialise in research practice. With this study, we give first empirical evidence on the use of LLMs in the research process. To this end, we examined around 20 use cases.
In this paper, we focus on tasks related to programming specifically, as motivated by the context of scientific work.
Accordingly, we focus on three prevalent use cases that involve coding, namely (i) writing programs, e.g., for software prototypes,
(ii) data analysis, and
(iii) data visualization.
For those use cases, we explored to which degree they are supported by a number of current LLM-based tools, and report our findings.
In addition, we provide an outlook on the broader set of use cases. Following open science principles, and to support replicability, we publish a replication package, recording all our interactions with the tools and our evaluation criteria.\footnote{\url{https://github.com/luuca78/LLMs4Science}\label{fn:repo}}

The remainder of the paper is structured accordingly.
After discussing related work in \Cref{sec:related}, we present the methodology and results of the coding scenarios in \Cref{sec:main}.
Finally, we provide the outlook on the larger picture of LLMs for science in \Cref{sec:outlook}, and close with a summary in \Cref{sec:summary}.
\section{Current and Related Research}
\label{sec:related}
The potential of using LLMs in aiding the research process is currently a highly discussed topic. 
In an editorial, Susarla et al.\cite{Susarla:2023} explore the potential of LLMs in information systems research. They explore research question formulation, data collection, data analysis, and writing---as tasks in which LLMs could add benefit. 
Kasneci et\ al.\cite{kasneci2023chatgpt} discuss the opportunities and challenges of using LLMs for education. 
They also briefly touch on the potential of LLMs to assist in the writing process of research papers, by generating summaries or outlines, and providing guidance on orthography.
This discussion also takes place across disciplines. For example, Grossmann et\ al.\cite{grossmann2023ai} share their perspective on LLM's transformative use for social science research practices and Li et\ al.\cite{li2023ethics} comment on the ethical concerns related to their use in medicine and medical research.
In their global survey on postdocs, \textit{Nature} found that 31\% of postdocs are already using AI-assisted chatbots like ChatGPT. Of those who use it, the two most reported use cases are text refinement and generating or troubleshooting code.\cite{le2023chatgpt} 

Beyond proprietary models, several open source LLMs exist, such as \textit{GPT-Neo}\cite{black2022gpt} or \textit{GPT-J}.\footnote{\textit{GPT-J}, \url{https://github.com/kingoflolz/mesh-transformer-jax}, accessed 2023-10-23.} There is also work focusing on training LLMs specifically for the use in a research context. 
\textit{SciBert}\cite{beltagy2019scibert} and 
\textit{ScholarBert}\cite{hong2023diminishing} are both pre-trained language models based on a corpus of scientific publications. These models are evaluated based on their performance executing automated tasks related to scientific practice, such as classification or relation extraction.
\textit{Galactica},\cite{taylor2022galactica} another large language model trained on scientific corpi, is a cautionary tale. Galactica was accessible online and meant to assist researchers in performing research tasks. The tool was taken down only days after its publication, following criticism over its tendency to hallucinate and confabulate.\cite{technologyreviewMetasLatest} This highlights the need for rigorous research and evaluation practices. The search for comprehensive evaluation systems for LLMs is ongoing research.\cite{chang2023survey}

There exists also a growing body of work on LLMs trained on code. However, we have to note here that LLMs of code is an emergent research field, and a review in this space inadvertently involves a high amount of grey literature. While we carefully selected these, it is essential to exercise caution with non-peer-reviewed studies.
The arguably most widely deployed proprietary model is \textit{Codex}, as it is integrated within GitHub Copilot.\cite{chen2021evaluating} Open source alternatives such as \textit{CodeParrot}\footnote{\textit{CodeParrot}, \url{https://huggingface.co/codeparrot/codeparrot}, accessed 2023-10-23.}, \textit{CodeBert}\cite{feng2020codebert}, or \textit{PolyCoder}\cite{xu2022systematic} exist; however, their functional correctness with regard to code completion tasks is significantly lower compared to Codex.\cite{xu2022systematic} To assess functional correctness of code generation, Chen et\ al.\ present an evaluation framework: \textit{HumanEval}.\cite{chen2021evaluating} HumanEval includes a Python data set of hand-written programming problems and several unit tests. Generated code is considered correct when it passes the associated unit tests. The fraction of passed samples is reported as a quantitative measure. Other quantitative evaluation frameworks have since then proposed, such as \textit{MBPP}\cite{austin2021program} or \textit{DS-1000}\cite{lai2023ds}. The latter specifically targeting code generation for data science. These frameworks are used in multiple studies investigating the functional correctness.\cite{xu2022systematic, ni2023l2ceval} 
Beyond correctness, Ouyang et\ al.\cite{ouyang2023llm} study the high degree of non-determinism in ChatGPT and Huang et\ al.\cite{huang2023bias} present an assessment framework to detect social biases in code generation. Vaithilingam et\ al.\cite{vaithilingam2022expectation} perform a user study to investigate how GitHub Copilot affects programming task completion times compared to a regular code completion tool. Their findings are surprising: ``Copilot did not necessarily reduce the task completion time or increase the success rate of solving programming tasks in a real-world setting.''\cite{vaithilingam2022expectation} Beyond quantitative metrics, Vaithilingam et\ al.\ find that study participants still prefer to use Copilot over a regular code completion tool. Conversely, Peng et\ al.\cite{peng2023impact} report productivity gains using Copilot of around 50\%.
This highlights the need for research investigating factors besides functional correctness. 

We present a use case-based evaluation, across a wide range of proprietary tools. Our evaluation assesses quantitative and qualitative aspects of the generated code. While our use cases are seemingly simple, meaningful differences in how they are handled by different tools are discovered. Our results also highlight the capabilities of this class of tools at large.

\section{LLM Tools for Code Generation}
\label{sec:main}

\subsection{Methodology}
\label{sec:method}

We selected a wide range of available LLM-based tools: \textit{ChatGPT with GPT 3.5}, \textit{ChatGPT with GPT 4}, \textit{Google Bard}, \textit{Bing Chat}, \textit{YouChat}, \textit{GitHub Copilot}, and \textit{GitLab Duo}.\footnote{See \url{https://chat.openai.com}, \url{https://bard.google.com}, \url{https://www.bing.com/new}, \url{https://about.you.com/youchat/}, \url{https://github.com/features/copilot}, \url{https://about.gitlab.com/gitlab-duo/}, all accessed 2023-11-10. For ChatGPT with GPT 3.5 and ChatGPT with GPT 4, we will write \textit{GPT 3.5} and \textit{GPT 4}, respectively. We conducted our experiments with the tools from 2023-10-24 till 2023-11-10. }
ChatGPT and Bing Chat are based on the generative pre-trained transformer (GPT).\cite{floridiGPT} Google Bard is based on \textit{PaLM}\cite{chowdhery2022palm}, the specifics of YouChat are not known. 
GitHub Copilot is based on \textit{Codex}, a GPT model trained on GitHub repositories.\cite{chen2021evaluating} GitLab Duo is based on \textit{Claude 2}, a proprietary model offered by \textit{Anthropic}.\footnote{\url{https://www.anthropic.com/index/claude-2}, accessed 2023-11-10}
To assess their capabilities in assisting researchers with coding tasks, we selected the following use cases.
\begin{enumerate}
    \item \textbf{Code generation}: Matrix multiplication in Java, using multi-threading. A complex aspect of object-oriented programming but which can be solved in a succinct manner.
    \item \textbf{Data analysis}: Given some data, generate Python code to analyze the data provided through different tasks and questions.
    \item \textbf{Data visualization}: Given some data, generate R code to visualise it in different ways.
\end{enumerate}
\noindent The full prompts can be found in our replication package (see \Cref{fn:repo}). For each use case, we used two variants of prompts, which were input independently by the authors of this paper. For the data analysis and visualisation cases, we used multiple consecutive prompts, each building on the context of the previous one.
We then assessed the generated code based on the following general criteria.
With \textit{Correctness}, we report whether the code produces correct results, and whether human intervention was required to obtain correct code -- that is, when the initial generated code was incorrect, we asked the tools to correct the errors. With \textit{Efficiency}, we report on run time performance. For data analysis and visualisation tasks, we also report on the quality of the analysis and visualisation. With \textit{Comprehension}, we give a measure on how hard or easy the code is to comprehend. Here, we focus on the quality of comments and documentation the tool generated. We also report on the length of the generated code, as we observed significant differences across different tools.
The assessment was done using an assessment rubric (see our replication package in \Cref{fn:repo}), with well defined criteria, and in discussions among all authors.
\subsection{Results}
\subsubsection{Code Generation}
\begin{center}
\begin{table*}[]
\scriptsize
\caption{Results for the matrix multiplication use case. Based on different criteria, we provide an overall qualitative rating (see our replication package for more detail) from one ($\bigstar$) to five stars ($\bigstar$$\bigstar$$\bigstar$$\bigstar$$\bigstar$).}
\centering
\begin{tabularx}{\textwidth}{ X l c c c c c c c }
\toprule
\multicolumn{2}{l}{\textbf{Matrix Multiplication}} & GPT-3.5 & GPT-4.0 & 
\begin{tabular}[x] {@{}c@{}}Bing\\Chat\end{tabular} & \begin{tabular}[x] {@{}c@{}}Google\\Bard\end{tabular} & YouChat & \begin{tabular}[x]{@{}c@{}}GitHub\\Copilot\end{tabular} & \begin{tabular}[x] {@{}c@{}}GitLab\\Duo\end{tabular} 
\\
\midrule
\multirow{2}{*}[-.2em]{Correctness} 
 & Correct without Intervention & $\checkmark$ & $\checkmark$ & $\checkmark$ & $\times$ & $\checkmark$ & $\checkmark$ & - \\ \cmidrule{2-9}
& Handles Edge Cases & $\checkmark$ & $\checkmark\tnote{*}$ &  $\checkmark\tnote{*}$ & $\times$ & $\checkmark$ & $\times$ & - \\ 
\midrule
Efficiency
& Benchmark (ms) & 471 &
579 &
1,899 &
648 &
446 & 515 & -
\\ 
\midrule
\multirow{3}{*}[-.5em]
{Comprehensibility}
& Code Commentary & $\times$ & $\times$ & $\times$ & $\checkmark$ & $\times$ & $\times$ & - \\ \cmidrule{2-9}
& Helpful Documentation & $\checkmark$ & $\checkmark$ & $\times$ & $\checkmark$ & $\times$ & $\times$ & -  \\ \cmidrule{2-9}
& Succinct Code & $\times$ & $\times$ & $\checkmark$ & $\times$ & $\checkmark$ & $\checkmark$ & - \\ 
\midrule
\addlinespace[-1pt]
\midrule
\addlinespace[2pt]
Overall Rating & & $\bigstar\bigstar\bigstar\bigstar$ & $\bigstar\bigstar\bigstar$ & $\bigstar\bigstar\bigstar$ & $\bigstar\bigstar\bigstar$ & $\bigstar\bigstar\bigstar\bigstar$ & $\bigstar\bigstar\bigstar$ & 
\\
\bottomrule
\addlinespace[2pt]
\multicolumn{9}{l}{\footnotesize\tnote{*} With the exception of the empty matrices edge case.} \\
\end{tabularx}
\label{tab:matrix}
\end{table*}
\end{center}
Table~\ref{tab:matrix} summarises our results for the matrix multiplication use case. Most tools  were able to generate correct executing code on first attempt. Google Bard required human intervention. More severely, GitLab Duo was only able to generate a single-threaded implementation, consequently, we were not able to rate it. We also tested a range of edge cases. Here again, most tools were able to handle all edge cases.\footnote{The edge cases comprised: empty matrices,
single element matrices (vector),
matrices with large numbers,
rectangular matrices (where one dimension is much larger than the other), matrices with negative numbers, identity matrices, and incompatible matrices.} We also benchmarked the generated code. We have conducted ten runs per tool. As input for each run, we randomly generated matrices sized 1000x1000. The reported result is the average of those ten runs. The runs were conducted on a consumer grade machine with an \textit{AMD Ryzen 7 5700U} processor with a clock rate of 1.8\,GHz, 8 cores, and 16\,GB of RAM. Additionally, we compared the generated code to a manually created single threaded reference implementation. Compared to that, all tools were able to achieve a significant speed up with their multi-threaded implementation. Here, YouChat generated the most performant code; while Bing Chat is notably slower than the other tools. When it comes to comprehension, GitHub Copilot, Bing Chat, and YouChat did not generate any commentary or documentation along the code. However, compared to other tools, they generated rather succinct code. 
With the exception of GitLab Duo, all tools generated solid results. Google Bard showed weaknesses with respect to correctness, and Bing Chat with respect to performance.
\subsubsection{Data Analysis and Data Visualisation}
\begin{table*}[]
\scriptsize
\caption{Results for the data analysis use case. Based on different criteria, we provide an overall qualitative rating (see our replication package for more detail) from one ($\bigstar$) to five stars ($\bigstar$$\bigstar$$\bigstar$$\bigstar$$\bigstar$).}
\begin{tabularx}{\textwidth}{ X l c c c c c }
\toprule
\multicolumn{2}{l}{\textbf{Data Analysis}} & GPT-3.5 & GPT-4.0 & \begin{tabular}[x] {@{}c@{}}Bing\\Chat\end{tabular} & \begin{tabular}[x] {@{}c@{}}Google\\Bard\end{tabular} & YouChat 
\\
\midrule
Correctness 
 & Correct without Intervention & $\times$ & $\checkmark$ & $\times$ & $\times$ & $\times$ \\ 
\midrule
\multirow{2}{18pt}[-.3em]{Efficiency}
& Appropriate \& Accurate Analysis & $\checkmark$ & $\checkmark$ & $\times$ & $\times$ & $\checkmark$
\\ \cmidrule{2-7}
& Benchmark (s)  & $< 1$ &
$< 1$ &
$< 1$ &
$< 1$ &
$< 1$ \\ 
\midrule
\multirow{3}{*}[-.5em]
{Comprehensibility}
& Code Commentary & $\checkmark$ & $\checkmark$ & $\checkmark$ & $\checkmark$ & $\times$
\\ \cmidrule{2-7}
& Helpful Documentation & $\checkmark$ & $\checkmark$ & $\checkmark$ & $\checkmark$ & $\checkmark$ \\ \cmidrule{2-7}
& Succinct Code & $\checkmark$ & $\times$ & $\checkmark$ & $\times$ & $\times$
\\
\midrule
\addlinespace[-1pt]
\midrule
Overall Rating & & $\bigstar\bigstar\bigstar\bigstar$ & $\bigstar\bigstar\bigstar\bigstar$ & $\bigstar\bigstar\bigstar$ & $\bigstar\bigstar$ & $\bigstar\bigstar\bigstar$ 
\\
\end{tabularx}
\label{tab:analysis}
\end{table*}
\begin{table*}[]
\scriptsize
\caption{Results for the data visualisation use case. Based on different criteria, we provide an overall qualitative rating (see our replication package for more detail) from one ($\bigstar$) to five stars ($\bigstar$$\bigstar$$\bigstar$$\bigstar$$\bigstar$).}
\centering
\begin{tabularx}{\textwidth}{ X l c c c c c c c }
\toprule
\multicolumn{2}{l}{\textbf{Data Visualisation}} & GPT-3.5 & GPT-4.0 & \begin{tabular}[x] {@{}c@{}}Bing\\Chat\end{tabular} & \begin{tabular}[x] {@{}c@{}}Google\\Bard\end{tabular} & YouChat 
\\
\midrule
Correctness 
 & Correct without Intervention & $\checkmark$ & $\checkmark$ & $\times$ & $\times$ & $\times$ \\ 
\midrule
\multirow{2}{18pt}[-.3em]{Efficiency}
& Quality of Generated Graph & $\bigstar\bigstar$ & $\bigstar\bigstar\bigstar\bigstar$ & $\bigstar$ & $\bigstar$ & $\bigstar$ 
\\ \cmidrule{2-7}
& Benchmark (s)  & $< 1$ &
$< 1$ &
$< 1$ &
$< 1$ &
$< 1$ \\ 
\midrule
\multirow{3}{*}[-.5em]
{Comprehensibility}
& Code Commentary & $\checkmark$ & $\checkmark$ & $\checkmark$ & $\checkmark$ & $\checkmark$
\\ \cmidrule{2-7}
& Helpful Documentation & $\checkmark$ & $\checkmark$ & $\checkmark$ & $\checkmark$ & $\checkmark$ \\ \cmidrule{2-7}
& Succinct Code & $\times$ & $\checkmark$ & $\times$ & $\times$ & $\checkmark$
\\
\midrule
\addlinespace[-1pt]
\midrule
\addlinespace[4pt]
Overall Rating & & $\bigstar\bigstar\bigstar\bigstar$ & $\bigstar\bigstar\bigstar\bigstar\bigstar$ & $\bigstar\bigstar$ & $\bigstar$ & $\bigstar\bigstar\bigstar$ 
\\
\end{tabularx}
\label{tab:data}
\label{tab:visualisation}
\end{table*}
Tables~\ref{tab:analysis} and~\ref{tab:visualisation} summarise our results for the data analysis and visualisation use cases, respectively. 
A specific challenge that we encountered in these use cases is the correct interpretation of data format and data type. For GitHub Copilot and GitLab Duo, we were not able to design a prompt that would lead to the desired results. 
We were not able to give them sufficient information about data format (e.g., table structure, cell names, ...) due to their focus on code completion, which makes interactive human intervention difficult. They only allow basic queries (A few words or at most one sentence) which specify the behaviour of a function.
While the other tools also produced data type mismatch errors, they were able to correct their errors after intervention.
A notable exception is GPT-4, which did not require any intervention for both use cases.
When it comes to the efficiency of the generated analysis, Bing Chat and Google Bard generated misleading results. For example, both ignored that we wanted to analyse profitability by department, not overall profitability.\footnote{We have compiled some examples of misleading results for both use cases at \url{https://github.com/luuca78/LLMs4Science}.}

For data visualisation, GPT-4.0 generated graphs of consistent good quality. GPT-3.5 and YouChat generated good graphs in the majority of cases. Bing Chat and Google Bard generated misleading visualisations. For example, generating a heatmap of performance by person, instead of performance by department, as asked.
Only GPT-4.0 included additional helpful indicators, such as regression lines for showcasing trends. 
As performance for data analysis and visualisation is not as critical as in production code, we only tested whether tools performed within a reasonable upper bound for our mostly small data sets. Overall, there was no tool which generated unreasonably slow code. Also, across most tools, generated commentary and documentation was of good quality. Overall, GPT-4.0 performed best for both cases. For data visualisation, it performed considerably better. 
Bing Chat and Google Bard had notable issues in generating accurate and appropriate analysis or visualisation, often producing misleading results and visuals.
Surprisingly, Google Bard was not able to produce running code for some of our visualisation prompts. 
\subsection{Limitations and Threats to Validity}
The research described above is early work, investigating a new subject of study, without established standards. The search for comprehensive evaluation systems for LLMs is ongoing research.\cite{chang2023survey} We proposed a method, but assume that over the coming years the research community will define and incrementally refine standards. A major issues regarding the assessment of LLMs revolve around non-determinism.\cite{ouyang2023llm} To mitigate associated threads, our results are based on two authors using the tools independently from each other with varying prompts. Our replication package contains full logs of each interaction.
Most of our qualitative assessments, such as the quality of the generated graphs, are subjective. To address this, we performed independent assessments based on an assessment rubric, with well defined criteria. 
\section{Outlook: LLMs for Science}\label{sec:outlook}
%
In the previous sections, we focused on use cases related to code generation. 
Here we provide a broader view on additional use cases relevant to research practice, albeit from a high level of abstraction due to space limitations.
The additional use cases we study are:
\begin{itemize}
    \item Text enhancement: detecting orthography errors or inconsistent terminology, suggestions on writing style and structure, expanding or shortening text;
    \item Generation of summaries and comparing introduction and conclusion text with the body of a paper;
    \item Finding and suggesting additional literature, based on a given paper or parts thereof;
    \item Conversational interface to literature, be it collections or individual publications;
    \item Providing reviewer-like feedback to authors;
    \item Checking reviews against reviewing guidelines;
    \item Generating presentation slides (including graphics) from papers, and generating paper drafts from presentation slides.
\end{itemize}
Across the investigated use cases, our early results underline the potential of these tools to aid in the scientific process -- specifically, in the process of writing. However, it is imperative to approach the outputs of these tools with caution. We repeatedly encountered instances of inaccuracies and confabulation, particularly in listed references.
It was also surprising to observe that some LLMs had significant difficulties in addressing supposedly simple tasks. For instance, Google Bard faced challenges in expanding the content to a specific word limit and expanded it by a factor of three, while YouChat exhibited a tendency to detect grammar or spelling errors where none existed. 
We aim to report on these case in more detail in future work.

\section{Summary, Discussion, and Future Work}\label{sec:summary}
LLM-tools are already used in scientific practice and beyond.
Given the substantial potential for increased productivity, we believe LLMs4Science and GenAI4Science are research topics of very high relevance. 
We elicited about 20 use cases in this category, and described our findings from assessing the three use cases related to programming.
While we studied seemingly simple use cases, results across tools were vastly different. With some tools generating efficient code while others not being able to produce code that compiles, even with human intervention. Overall most tools performed well for the matrix multiplication case. However, more creative and context-specific tasks, like performing data analysis and visualisation over multiple consecutive queries, yielded more diverse results. As such, some tools had difficulty to pick up on important details, which lead to wrong and misleading analysis results. 
This highlights the danger for the integrity of results. Detecting such misleading statements can be challenging, especially when the abilities of the tools exceed the skills of the person using them. For example, we encountered visuals were the graph was labeled in the expected way -- but closer investigation revealed that the wrong data was displayed. For more complex data, this can constitute a major challenge. The tendency to confabulate can substantially undermine the promised productivity gains, when results can not be relied upon.

In this paper, we focused on LLM-based tools, since we believe this is how most scientists would use such tools currently.
The last author of this paper introduced such an LLM-based tool, called \textit{FhGenie}, in June 2023 at Fraunhofer.\cite{fraunhoferFhGenieFraunhoferGesellschaft} Fraunhofer is a German research society with over 70 institutes and approx.\ 30,000 staff. FhGenie is actively used by several thousand researchers at the time of writing, and the (unstructured) feedback has been positive. At Fraunhofer, all users of FhGenie (and researchers in particular) are personally and solely responsible for their use of the outputs, primarily since confabulation cannot be prevented by any technical means in the current system. 
We believe this attribution of responsibility to be the right one for the foreseeable future, even when the chance of confabulation gets reduced.

In future work, we plan to study the use cases in more detail and develop our methodology further, applying it to larger collections of existing evaluation frameworks, such as HumanEval. We aim to investigate, to which degree our methodology can be supported by automation.
While we have presented a first set of criteria that goes beyond pure functional correctness, this is by no means exhaustive. Further dimensions of importance must be studied, such as the tendency of LLMs to perpetuate biases. Using such tools responsibly should be at the forefront of considerations for researchers and developers alike.

\bibliography{main}%

\end{document}